\documentclass[useAMS,usenatbib]{mn2e}
\pdfoutput=1
\usepackage{graphicx}
\usepackage{bmpsize}
\usepackage{multirow}
\usepackage{gensymb}
\usepackage{amsmath}
\usepackage{amssymb}
\usepackage{subfig}
\usepackage{rotating}
\usepackage{times}

\title[SFE after cloud envelope loss]{Can the removal of molecular cloud envelopes by external feedback affect the efficiency of star formation?}
\author[W.~E.~Lucas et al.]{William E.~Lucas$^1$\thanks{E-mail: 
  wel2@st-andrews.ac.uk}, Ian A. Bonnell$^1$ and Duncan H. Forgan$^1$\\
  $^1$ SUPA, School of Physics \& Astronomy, University of
    St Andrews, North Haugh, St Andrews, Fife KY16 9SS, United Kingdom}
  
\begin{document}

\maketitle

\begin{abstract}  
We investigate how star formation efficiency can be significantly decreased by the removal of a molecular cloud's envelope by feedback from an external source. Feedback from star formation has difficulties halting the process in dense gas but can easily remove the less dense and warmer envelopes where star formation does not occur. However, the envelopes can play an important role keeping their host clouds bound by deepening the gravitational potential and providing a constraining pressure boundary. We use numerical simulations to show that removal of the cloud envelopes results in all cases in a fall in the star formation efficiency (SFE). At $1.38$ free-fall times our $4\,\mathrm{pc}$ cloud simulation experienced a drop in the SFE from 16 to six percent, while our $5\,\mathrm{pc}$ cloud fell from 27 to 16 per cent. At the same time, our $3\,\mathrm{pc}$ cloud (the least bound) fell from an SFE of $5.67$ per cent to zero when the envelope was lost. The star formation efficiency per free-fall time varied from zero to $\approx 0.25$ according to $\alpha$, defined to be the ratio of the kinetic plus thermal to gravitational energy, and irrespective of the absolute star forming mass available. Furthermore the fall in SFE associated with the loss of the envelope is found to even occur at later times. We conclude that the SFE will always fall should a star forming cloud lose its envelope due to stellar feedback, with less bound clouds suffering the greatest decrease.
\end{abstract}

\section{Introduction} \label{s:intro}

Star formation is a complex process governed by several intertwined physical processes (\citealt{E99}; \citealt{DEA142}). The core mechanism is the collapse under gravity of dense regions of gas into stars, with this for the most part taking place in giant molecular clouds with only around ten per cent of stars forming in isolation (\citealt{LL03}; \citealt{EEA09}). Opposing this are thermal pressure along with turbulent kinetic motions (\citealt{HB04}; \citealt{MLK04}) and magnetic fields (\citealt{PB09}; \citealt{PN11}). The confluence of these processes governs the properties of a new stellar population such as the number of stars, their masses, locations, multiplicity and dynamics (\citealt*{SAL87}; \citealt*{BLZ07}).

One of the most important questions concerning star formation is why it is so inefficient, with of order a few percent of the gas available on large scales being transformed into stars (\citealt{KE12}; \citealt{KR14}; \citealt{HEA16}). Feedback from young stars, in the form of ionizing radiation, stellar winds or supernova explosions, is often invoked as a solution, but numerical simulations have repeatedly shown that feedback has at most a moderate effect on the star formation efficiency of dense gas (\citealt*{DBW07} and other papers in this series; \citealt{DEA14}; \citealt{MLEA15}). The effect is much more significant in lower density gas, but it is instead in the higher density gas that star formation takes place. This would seem to indicate that feedback cannot greatly impact the efficiency of star formation.

In reality star formation takes place in dense and cold molecular hydrogen gas at the centre of these clouds. The optical depth here is high enough that they are shielded from the external Galactic radiation field, while the outer layers of the cloud are exposed and form an envelope of atomic hydrogen (\citealt{AEA92}; \citealt{LBEA93}). Observations of star forming clouds have found them to typically be marginally bound, though the virial parameter for individual clouds varies from $0.1$ to $10$ (\citealt{BEA08}; \citealt*{HEA09}; \citealt{WEA11}; \citealt{DBP11}; \citealt{DEA142}).

Restricting a cloud rather than allowing it to freely expand into a vacuum would keep its average volume density higher than otherwise. \citet{KT07} argued using observations that a cloud's star formation efficiency (SFE) per free-fall time is independent of its density, allowing \citet*{KDM12}, \citet{F13} and \citet*{SFK15} to formulate star formation laws based on turbulence. Simulations by \citet*{CBK08} found that the SFE increases for more tightly bound clouds. Newer simulations by \citet*{BEA11} and observations by \citet{LEA14} found a positive correlation between SFE and density. \citet{EL77} introduced the concept of sequential star formation taking place as ionizing feedback from young stellar clusters repeatedly sweeps up material to form new clusters. As such we ask how the SFE might change with the loss of a cloud's envelope due to feedback in a similar scenario, though with the feedback here taking a negative rather than positive form.

Observations and models agree that the transition from warm low density gas to cold high density gas is gradual and that there is a mixture of phases present \citep{GL05}. To test the envelope's effect on star formation efficiency, we envision a simplified geometry with a warmer low density envelope surrounding the cold, dense and turbulent gas where star formation can occur. The extra gravity and pressure from the envelope keeps the super-virial cloud from expanding. Two scenarios follow: one in which the envelope remains in place, and one in which it is later removed as a first order approximation to its stripping by stellar feedback. In this paper we report on simulations of such a cloud within a transient warm envelope in order to ask whether such feedback can contribute to a significant reduction in the star formation efficiency in the dense gas.

\section{Numerical method} \label{s:method}

Smoothed particle hydrodynamics (SPH) is a Lagrangian technique used to model a fluid as a set of particles. Each particle's mass is spread throughout a volume with the use of a smoothing kernel. The fluid equations may then be reformulated as sums over the particles by treating the density, internal energy and so on as interpolated quantities. The simulations shown later in this paper were performed with the SPH code \textsc{sphNG} (\citealt*{BBP95}; \citealt{PM07}). This code allows for the creation of sink particles which represent stars. As a Lagrangian method, SPH also allows for mass to be traced throughout a simulation. This makes the method widely used in astrophysics, and as such many papers provide in-depth descriptions. Some recommendations are \citet{BE90} and \citet{MO92} for basic overviews, and \citet{PM04} for details of the grad-h formalism which was used here.

The basic simulation setup is as follows. Firstly, a spherical cloud of radius $5 \, \mathrm{pc}$ and mass $10^4 M_\odot$ was created. The original density was uniform with a value of $\rho_0 = 1.29 \times 10^{-21} \,\mathrm{g}\,\mathrm{cm}^{-3}$, giving a free-fall time of $1.85 \times 10^6 \,\mathrm{years}$. Its temperature was set to $10\,\mathrm{K}$. The region from $5$ to $10\,\mathrm{pc}$ was then filled with gas at $10^3\,\mathrm{K}$ and a density a factor of ten lower then the inner cloud. The sound speed in the cold inner cloud was $0.253\,\mathrm{km}\,\mathrm{s}^{-1}$ while in the warm outer envelope it was $2.53\,\mathrm{km}\,\mathrm{s}^{-1}$. Note that the gas in both structures has a mean molecular weight of $1.29$, that for a mix of atomic hydrogen and helium, and differs only in temperature. Allowing it to change with phase would be desirable for a `complete' simulation, but here it is kept constant both as a numerical convenience and in order to keep the model simple. From this point the terms `cloud' and `envelope' will refer to the inner and outer structures respectively.

\begin{figure*}
	\includegraphics[width=17.5cm]{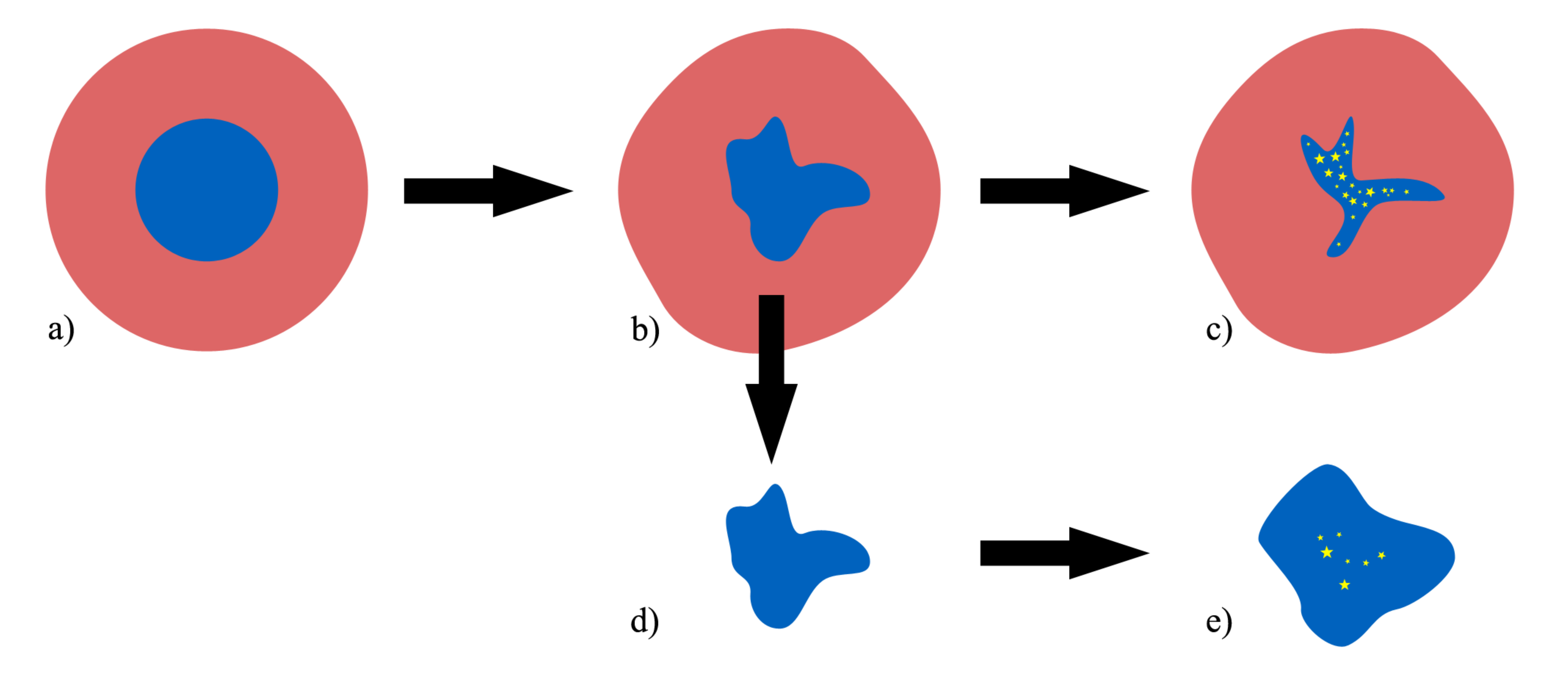}
	\caption{Simulation method. In subfigure a), a cold turbulent cloud (blue) is embedded within a warm nonturbulent envelope (red). This was then evolved forwards in time, shown in subfigures b) and then c). Part of the way through the simulation, usually at a quarter of the turbulent crossing time and here shown in b), a separate simulation was created by removing the warm envelope to create the setup seen in subfigure d). The cloud was then evolved in isolation to a later time shown in e) to allow a direct comparison with the original run in which the envelope was retained for the entire duration of the simulation. It was expected that the loss of the envelope would lead to more star formation taking place in c) than in e). (A colour version of this plot is available in the online version.)}
	\label{fig:evolution_method}
\end{figure*}

A turbulent velocity field following an approximately Kolmogorov power spectrum of $\langle | v_k |^2 \rangle \propto k^{-3.5}$ was applied to the cloud. The $\alpha$ value \citep{BB79} is the ratio of the kinetic plus thermal energy to the gravitational binding energy. The velocities were scaled such that, when considering the cloud in isolation, it would be unbound with $\alpha = 2.00$. As such the root mean square (RMS) speed was $4.53\,\mathrm{km}\,\mathrm{s}^{-1}$ (Mach number $\mathcal{M} = 17.9$), a value slightly above the $3$ to $4 \,\mathrm{km}\,\mathrm{s}^{-1}$ that might be expected for a cloud of this size and mass from the Larson relations (\citealt{L81}; \citealt{REA03}). 

The envelope on the other hand was allowed to remain with no internal motions. Either thermal or ram pressure from the envelope could have been used to constrain the cloud, so while it was set to a temperature of $10^3\,\mathrm{K}$, a lower value could have been used and the envelope also made turbulent. For simplicity, we used thermal pressure alone.

The same simulation was run with a less bound cloud by taking the original cloud of radius $5\,\mathrm{pc}$ and removing the layer from $4$ to $5\,\mathrm{pc}$, then recreating the envelope by populating the region from $4$ to $10\,\mathrm{pc}$ with low density warm gas as before. This left the cold cloud with a mass of $5120 M_\odot$, and as such in isolation $\alpha = 3.15$. The $4$ and $5\,\mathrm{pc}$ setups were created at four resolution levels of $0.04 M_\odot$, $0.02 M_\odot$, $0.01 M_\odot$, and $2 \times 10^{-3} M_\odot$ per particle. Due to the initial random particle placement, the total cloud mass did vary very slightly between different resolution simulations when deriving the $4 \, \mathrm{pc}$ clouds. A single simulation was similarly set up to run with a cloud size of $3 \,\mathrm{pc}$ at the $0.01 M_\odot$ resolution level. With a mass of $2160 M_\odot$, this cloud was most unbound with $\alpha = 5.59$. 

\begin{table} \centering
	\caption{The simulation names and meanings. The name reflects the cloud radius in parsecs and the resolution of the simulation (very low, low, medium or high). The second column gives the gas particle mass in solar units. The final two columns provide the mass of the cloud and envelope. Each run had two variations -- one in which the warm envelope remained for the whole simulation's duration (indicated by prefixing the name given here with an `E') and another in which it was removed after a quarter of the cloud's crossing time (indicated by `X').}
	\begin{tabular}{@{}lccccc}
		\hline
		Run name & $m_\mathrm{part} / M_\odot$ & $M_\mathrm{cloud}  / M_\odot$ & $M_\mathrm{envelope}  / M_\odot$ \\
		\hline
		3pc--medres & $0.01$ & $2,160.72$ & $7,786.59$ \\
		4pc--vlowres & $0.04$ & $5,114.44$ & $7,479.88$ \\
		4pc--lowres & $0.02$ & $5,116.08$ & $7,482.26$ \\
		4pc--medres & $0.01$ & $5,119.46$ & $7,487.21$ \\
		4pc--highres & $2 \times 10^{-3}$ & $5,115.294$ & $7,481.118$ \\
		5pc--vlowres  & $0.04$ & $10,000.00$ & $7,000.00$ \\
		5pc--lowres  & $0.02$ & $10,000.00$ & $7,000.00$ \\
		5pc--medres & $0.01$ & $10,000.00$ & $7,000.00$ \\
		5pc--highres & $2 \times 10^{-3}$ & $10,000.00$ & $7,000.00$ \\
		\hline
	\end{tabular}
	\label{tab:sim_names2}
\end{table}

Dynamic sink particle creation was enabled in all simulations. The criteria for creation are as described in \citet{BBP95}; a brief description follows here. A gas particle being tested had to exceed a critical density set to $10^8 \rho_0 = 1.29 \times 10^{-13} \,\mathrm{g}\,\mathrm{cm}^{-3}$, and all its neighbours (gas particles within the kernel) had to be within a distance of $0.01 \,\mathrm{pc}$. This clump of particles also had to be bound and collapsing. The SPH code aimed to give each particle around 50 neighbours and so the minimum sink mass was $2 M_\odot$ in the `vlowres' simulations, $1 M_\odot$ in `lowres' simulations, $0.5 M_\odot$ in `medres' runs, and $0.1 M_\odot$ in `highres'. After sink creation, gas particles within the same distance of $0.01\,\mathrm{pc}$ could be accreted if it they passed a boundedness test. Particles within $2.5 \times 10^{-3} \,\mathrm{pc}$ were accreted without testing.

After setup, the simulations were evolved isothermally under self-gravity for as long as possible before being halted by the small timesteps required in the high density regions forming sinks. For each, split simulations were created after a quarter of the turbulent crossing time in the cloud (evaluating to $0.32\,\mathrm{Myr}$ for the $3\,\mathrm{pc}$ cloud, $0.43\,\mathrm{Myr}$ for the $4\,\mathrm{pc}$ cloud and $0.54\,\mathrm{Myr}$ for the $5\,\mathrm{pc}$ cloud) in which the entirety of the warm envelope was removed as an approximation to the effects of feedback. In a later set of simulations the envelope was removed from the medium resolution $4\,\mathrm{pc}$ cloud at later times of up to $1.50 t_\mathrm{ff}$. The process of evolution and creation of a parallel simulation without the envelope is shown in Figure~\ref{fig:evolution_method}.

The simulations and naming conventions are given in Table~\ref{tab:sim_names2}. The cloud radius in parsecs is provided, along with an indication of the resolution level of the simulation: very low, low, medium or high. When referring to a specific simulation, an `E' is prefixed to this to indicate that the envelope remained throughout the whole simulation, while an `X' prefix indicates that it was removed during the evolution.

\section{Evolutionary overview}

%

\begin{figure*}
	\includegraphics[width=17.5cm]{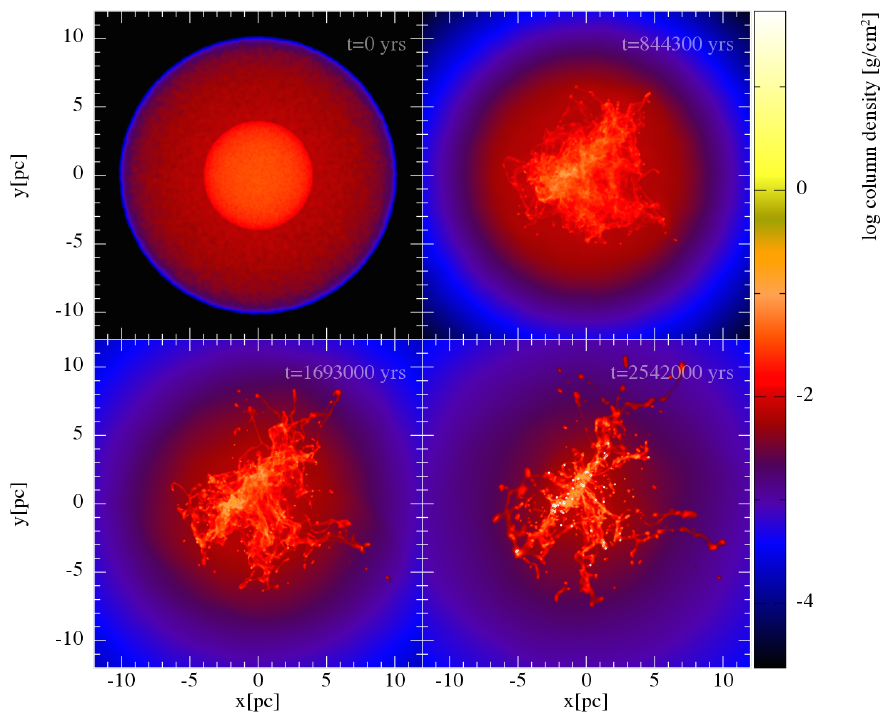}
	\caption{Evolution of simulation E4pc--medres. Shown are the column densities and sink particles (as white points) at four times. For reference, $844300\,\mathrm{years} = 0.46t_\mathrm{ff}$, $1693000\,\mathrm{years} = 0.92 t_\mathrm{ff}$ and $2542000\,\mathrm{years} = 1.38 t_\mathrm{ff}$. Turbulence in the cloud generated structure, with sink particles forming in the densest regions by the final time shown. Enough kinetic energy had been placed within the initial cloud to launch small blobs outwards, but with the presence of the envelope the cloud on the whole remained roughly the same size throughout. (A colour version of this plot is available in the online version.)}
	\label{fig:coldens-r4pc-evo}
\end{figure*}

\begin{figure*}
	\includegraphics[width=17.5cm]{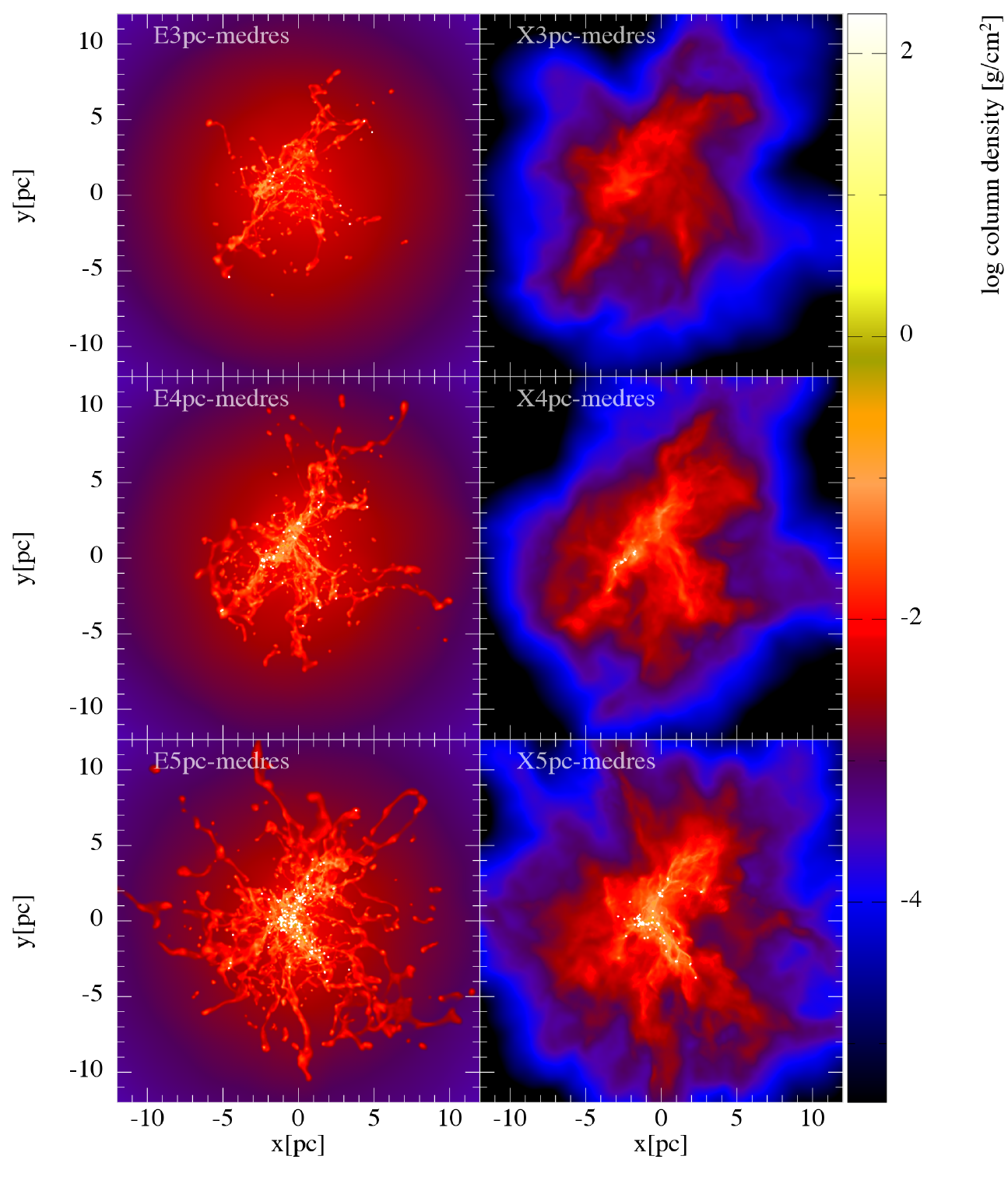}
	\caption{Column densities for all medium resolution simulations at $1.38t_\mathrm{ff} = 2.54 \,\mathrm{Myr}$, with dots showing the locations of sink particles. The plots are arranged with increasing cloud radius from top to bottom, and with the left-hand column showing the E- simulations which retained the envelope while in the X- simulations shown on the right the envelope was removed after a quarter of a crossing time. Gas reached higher densities in the E- simulations and formed more complex structure, while the cloud expanded to become much more diffuse in the X- simulations. Sink formation is generally associated with the densest regions. Increasing structure, density and star formation can also be seen as the cloud radius (and thus boundedness) increased. (A colour version of this plot is available in the online version.)}
	\label{fig:coldens_medres}
\end{figure*}

As might be expected, the cold inner cloud's evolution was characterised by the growth of overdensities throughout its evolution. These initially came about through the generation of structure via turbulence. Self-gravity was then able to take hold in the densest regions and bring them to the point where stars could be formed. 

The warm envelope, retained in the E- simulations, maintained a roughly spherical symmetry as it gained little kinetic energy from the turbulent central cloud, and thus lacked significant internal motions. In contrast the cloud itself developed strong filamentary structure along with a number of sub-parsec scale clumps containing dynamically forming sink particles. As the cloud was by design unbound, it expanded and partially displaced the envelope.

The X- simulations, in which the envelope was removed, exhibit quite different evolution. As was noted in Section~\ref{s:method}, they were removed at $0.25t_\mathrm{cross}$ once the turbulence had already started to generate the internal filamentary structure. In the absence of the envelope the cold cloud was free to expand. The filaments formed from the turbulence were more diffuse in comparison with the simulations in which the envelope was present throughout. This lower density translates to a decrease in the importance of self-gravity and hence a decrease in the local star formation rate. As such parallels in structure can be seen with the corresponding E- runs, but the small dense clumps seen previously did not form in these simulations.

Figure~\ref{fig:coldens_medres} shows column densities and sink locations for the medium resolution simulations at all three cloud sizes (3, 4 and $5\,\mathrm{pc}$) and both with and without the envelope, at $1.38t_\mathrm{ff} = 2.54 \,\mathrm{Myr}$. Across the three cloud sizes, the most apparent change is the correlation of the original cloud size with the extent of dense structured gas after evolution. Much of this takes the form of cometary blobs of cold gas launched outwards through the envelope. It is also immediately apparent that with the removal of the envelope in the E- runs, the ability of the cloud to form dense structure in general along with sink particles dropped dramatically. At the same time, while the $5\,\mathrm{pc}$ cloud (the most bound of the set) was able to form stars in the absence of the envelope, the $3\,\mathrm{pc}$ cloud had failed to form any. Indeed no sinks had been formed even by the end of E3pc--medres at $2.54 t_\mathrm{ff} = 4.71\,\mathrm{Myr}$. E4pc--medres and X4pc--medres lay between the extrema of the $3$ and $5\,\mathrm{pc}$ runs.

\begin{figure} \centering
	\includegraphics[width=9cm]{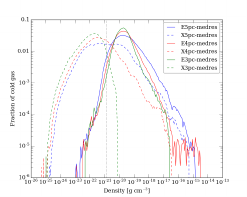}
	\caption{Mass-weighted density PDF of cold (10 K) gas in the 3pc--medres, 4pc--medres and 5pc--medres simulations at $1.38t_\mathrm{ff} = 2.54 \,\mathrm{Myr}$. The dotted vertical line shows the original mean density $\rho_0 = 1.29 \times 10^{-21} \,\mathrm{g}\,\mathrm{cm}^{-3}$ in the cold gas. Aside from the difference in densities between the three sets caused by the increase of mass with cloud size, it is apparent that densities were enhanced when the warm envelope remained in place throughout the simulation. (A colour version of this plot is available in the online version.)}
	\label{fig:density_pdf}
\end{figure}

Examining the density PDFs reveals in more detail the quantitative differences between the simulations. These are shown in Figure~\ref{fig:density_pdf} for the cold gas in the E- and X- variations of the 3pc--medres, 4pc--medres and 5pc--medres simulations at $1.38t_\mathrm{ff}$.

The density PDFs for both variations of the 3pc--medres and 4pc--medres runs as well as E5pc--medres followed approximately lognormal distributions. The distributions for the E- runs were located at higher densities, with excesses at the very highest densities representing the self-gravitating gas moving towards sink formation. This noticeably does not occur in X3pc--medres. In terms of shape the density PDF for X5pc--medres is the exception, seemingly spreading between the E- and X- groups. It is clear from examination of the column densities in Figure~\ref{fig:coldens_medres} that the removal of the warm envelope led to the cloud's expansion into the vacuum. In X5pc--medres, the same expansion is seen in gas moving to lower densities -- at the same time, the cloud in this simulation was closer to being globally bound, and examination of the simulations does show that the central region was undergoing collapse. This resulted in the `flat-top' of the density PDF for X5pc--medres, representing the dense inner regions of the cloud as well as a new envelope formed from the expansion of its outer layers.

At the time of the PDFs shown in Figure~\ref{fig:density_pdf}, the median density in E3pc--medres was $1.97 \times 10^{-20} \,\mathrm{g}\,\mathrm{cm}^{-3}$, while removing the envelope led to the median density in X3pc--medres falling to $1.42 \times 10^{-22} \,\mathrm{g}\,\mathrm{cm}^{-3}$. In E4pc--medres it was $2.19 \times 10^{-20} \,\mathrm{g}\,\mathrm{cm}^{-3}$, while it similarly fell in X4pc--medres to $4.00 \times 10^{-22} \,\mathrm{g}\,\mathrm{cm}^{-3}$. Finally the respective values for E5pc--medres and X5pc--medres were $2.58 \times 10^{-20} \,\mathrm{g}\,\mathrm{cm}^{-3}$ and $1.29 \times 10^{-21} \,\mathrm{g}\,\mathrm{cm}^{-3}$. 

Two observations may be made here. Firstly, changing the size of the cloud from $3$ to $5\,\mathrm{pc}$ led to only a small change in the median density in the E- simulations, while in the X- simulations it fell by almost a factor of ten. Secondly, removing the envelope dropped the median density by a factor of $20$ in the 5pc--medres simulations but a factor of $55$ for the 4pc--medres runs and $139$ for the 3pc--medres runs. That is to say, the density in the cloud was increased more by retaining the envelope than by making it larger and more bound. Thus we can say that the effect of the warm envelope contributed greatly towards maintaining higher densities in the cloud.

Approaching $1 t_\mathrm{ff}$, sink particles began to form in all simulations with the exception of X3pc--medres, concentrated in the cold and dense filaments and clumps. In no simulation did envelope particles take part in accretion to sink particles. Over time the proto-clusters coalesced to form larger structures following a process similar to that seen in \citet*{BBV03}, the largest forming from the dense central regions of the cloud that can be seen just above the origin in Figure~\ref{fig:coldens_medres}. Isolated sinks also formed from the smaller clumps at larger distances from the origin, but due to the clumps' absence, at a much lower rate in the X- simulations in which the envelope had been removed. 

An isothermal equation of state was used in all the simulations. Because of this, the gas may easily fragment as seen in \citet*{DBC05} and \citet{LA05} and the number of sinks formed is likely an upper limit, with the true value being lower. However the simulations should still provide a good estimate of the total mass in stars, and it is with this quantity we investigate the star formation efficiency in the next section.

\section{Star formation efficiency}
The star formation efficiency (SFE) was found to be very different between the various simulations. The SFE was defined in relation to the mass of cold ($10\,\mathrm{K}$) gas in the cloud, $M_\mathrm{cloud}$, a value differing between the variously-sized clouds used. It may then be simply calculated at any time by
\begin{equation}
	SFE = \frac{M_\mathrm{sinks}}{M_\mathrm{cloud} + M_\mathrm{sinks}},
\end{equation}
where $M_\mathrm{sinks}$ is the total mass in sink particles. As no envelope particles went into any sinks, the sum on the bottom always equalled the original cloud mass. Although the sink particles are only numerical representations of stars, they are a suitable proxy to allow the calculation of the SFE: while each one may not truly represent a star, it does still contain gas that was determined to be dense and gravitationally bound. As such, the SFEs reported here do represent the star forming capability of the cloud, within the limits of the resolution, which is covered in Section~\ref{ss:resolution}.

\subsection{Original simulations} \label{ss:sfe_orig}
\begin{figure} \centering
	\includegraphics[width=9cm]{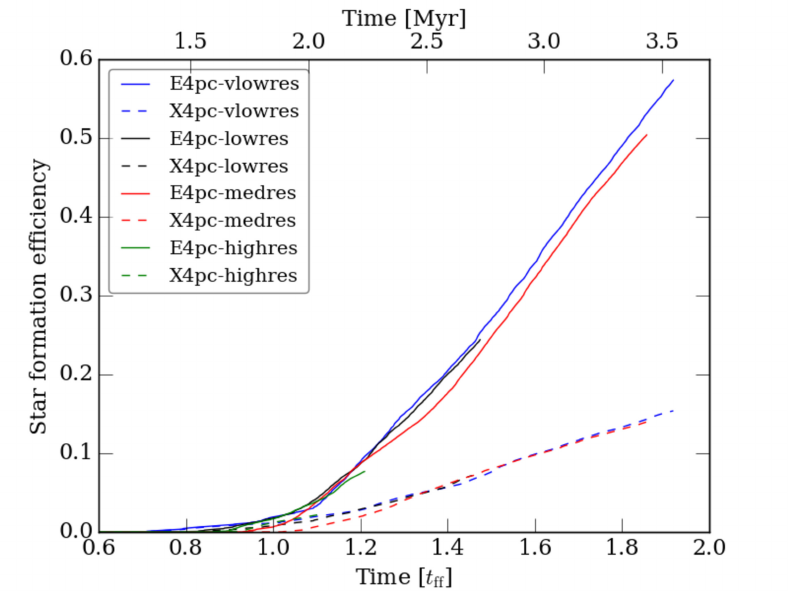}
	\caption{Star formation efficiency for the $4 \,\mathrm{pc}$ radius clouds. The low resolution simulations actually began forming sink particles earlier than the medium resolution runs, but at such low rates that the SFE was of order $~0.01$. Once the SFE reached an appreciable fraction of unity, the runs at various resolutions had essentially diverged to give two groups: the E- runs in which the envelope remained in place throughout, and the X- in which it was removed. By the endpoints of these runs, the E- simulations had an SFE a factor of almost four greater. (A colour version of this plot is available in the online version.)}
	\label{fig:sfe_r4pc_linlin}
\end{figure}

\begin{figure} \centering
	\includegraphics[width=9cm]{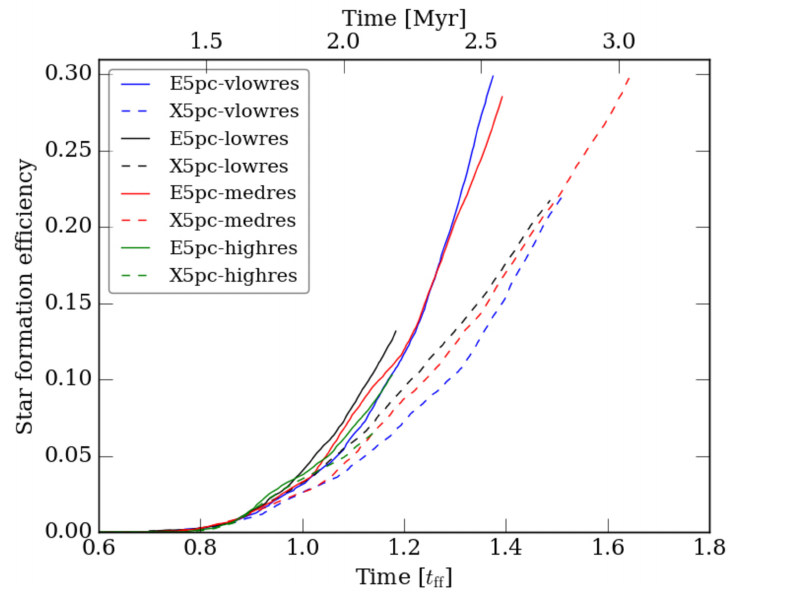}
	\caption{Star formation efficiency for the $5 \,\mathrm{pc}$ radius clouds, plotted in the same way as Figure~\ref{fig:sfe_r4pc_linlin}. At earlier times the separation between the E- and the X-runs is not as clear as it was with the $4\,\mathrm{pc}$ cloud, but as the simulation progressed the divergence between the two sets grew so that the increased SFE with the envelope's presence is clearly visible. (A colour version of this plot is available in the online version.)}
	\label{fig:sfe_r5pc_linlin}
\end{figure}

The SFEs of the $4\,\mathrm{pc}$ clouds are plotted against time in Figure~\ref{fig:sfe_r4pc_linlin} for both the E- and X- variants and at all four resolutions used for that setup. A similar plot is shown for the $5\,\mathrm{pc}$ clouds in Figure~\ref{fig:sfe_r5pc_linlin}. 

The profiles of the SFE for the two cloud sizes shown in Figures~\ref{fig:sfe_r4pc_linlin} and \ref{fig:sfe_r5pc_linlin} both follow a similar pattern. Before $\approx 1.0 t_\mathrm{ff}$ star formation proceeded but very slowly. After that point the SFE increased much more quickly to sizeable fractions of unity, with the efficiencies for the E- and X- simulations beginning to diverge shortly afterwards at $\approx 1.1 t_\mathrm{ff}$. After this point the two sets of SFEs continued to increase at roughly linear rates. The divergence appears more noticeable for the $4\,\mathrm{pc}$ cloud thanks to the longer runtimes that could be achieved for these simulations.

It is clear from both figures that the X- runs in which the warm envelope was removed were less efficient at forming stars when compared to the E- simulations in which the envelope remained throughout the simulation. Taking a fiducial point at $1.38 t_\mathrm{ff}$, the SFE in E4pc--medres was $0.16$ while it fell to $0.06$ in X4pc--medres. At the same time for the $5\,\mathrm{pc}$, E5pc--medres had an SFE of $0.27$ and X5pc--medres had $0.16$. By the latest times available for the $4\,\mathrm{pc}$ cloud, after $1.8t_\mathrm{ff}$, the difference was even greater, with the $\mathrm{SFE} \approx 0.15$ in the X- runs increasing to over $0.5$ in the E- runs.  

Though the values are too low to be seen in the figures, star formation began in the $4\,\mathrm{pc}$ cloud at different times, from $0.7$ to just over $1 t_\mathrm{ff}$ depending on the level of resolution and whether the envelope was present or not. Conversely, star formation in the $5\,\mathrm{pc}$ cloud began between $0.7$ and $0.8 t_\mathrm{ff}$ in all cases. An observation may be made in that all the simulations in which the envelope was retained began to form sink particles earlier than the variants in which the envelope was lost. Furthermore, the discrepancy was greater for the $4\,\mathrm{pc}$ clouds (where sink formation was delayed by up to $\approx 0.1 t_\mathrm{ff}$) than the $5\,\mathrm{pc}$ clouds (where it was $\approx 0.01 t_\mathrm{ff}$). Since the SFE at these times was generally of order $10^{-3}$, the end effect is negligible.

The $3\,\mathrm{pc}$ cloud is not shown on these plots, being much more extreme than the others. In E3pc--medres, by $1.38t_\mathrm{ff} = 2.54 \,\mathrm{Myr}$, 22 sinks had formed with an SFE of $5.67 \times 10^{-2}$. By the end of the simulation at $2.55 t_\mathrm{ff} = 4.71 \times 10^6 \,\mathrm{Myr}$, there were 103 sinks with an SFE of $0.59$. In contrast to all other runs, no sinks formed at all right up until the same end point for the simulation. The lower densities in the cloud that resulted from the envelope's removal, seen in the density PDFs in Figure~\ref{fig:density_pdf}, led to the complete inability of the cloud to form stars at all.

\subsection{Resolution effects} \label{ss:resolution}
It can be seen in Figures~\ref{fig:sfe_r4pc_linlin} and \ref{fig:sfe_r5pc_linlin} that although there were variations between the star formation efficiencies for simulations of different resolution levels, on the whole they followed one another closely and allowed us to confirm the effect of the envelope. This can in particular be seen with the longer runtimes for the $4\,\mathrm{pc}$ cloud simulations.

As noted in Section~\ref{ss:sfe_orig}, the $4\,\mathrm{pc}$ cloud in particular began to form sink particles at different times depending on the level of resolution. We believe that this is due to the lower levels being unable to resolve the turbulence imposed in the initial conditions, causing more energy to be deposited in larger scale motions and making it easier to form sinks on small scales. These variations were so small however that they had no real bearing on the ultimate outcome.

\subsection{Variation of the star formation efficiency with boundedness}

\begin{figure} \center
	\includegraphics[width=9cm]{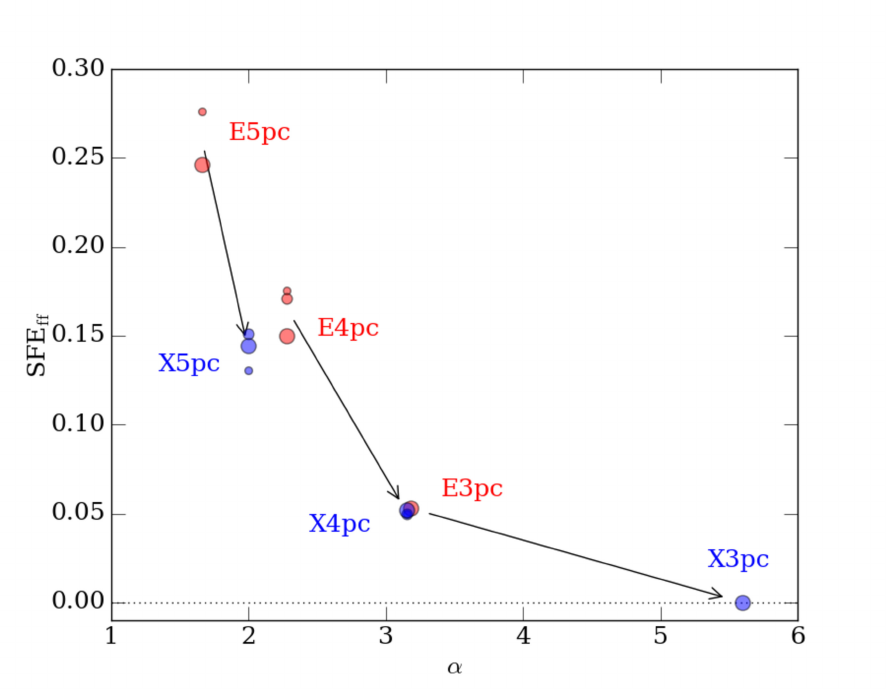}
	\caption{The star formation efficiency per free-fall time, $\mathrm{SFE}_\mathrm{ff}$, calculated at $1.38 t_\mathrm{ff}$ for all the simulations which had data as late as that, plotted here against $\alpha = \mathcal{T}/|W|$. These are both dimensionless quantities. The arrows show the change when removing the envelope from a cloud of a given radius. The simulation resolution is reflected by point size; the largest points here are the medium resolution simulations. It can be seen that for each cloud of a given radius, the $\mathrm{SFE}_\mathrm{ff}$ fell when the envelope was removed. At the same time the increase in $\alpha$ brings the two data sets roughly in line with one another. The point for E3pc--medres lies almost on top of those for X4pc--vlowres, --lowres and --medres. (A colour version of this plot is available in the online version.)}
	\label{fig:sfeff_vs_alpha}
\end{figure}

The ability of a cloud to hold itself together can be quantified by considering the $\alpha$ parameter (\citealt{BB79}; \citealt{T80}). We may take the sum of the cloud's kinetic and thermal energy to be the total energy opposing collapse $\mathcal{T}$. With the gravitational binding energy of the cloud $W$ the ratio 
\begin{equation} \label{eq:alpha}
	\alpha \equiv \frac{\mathcal{T}}{|W|}
\end{equation}
indicates the cloud's boundedness, with a value of $\alpha \le 1$ meaning that the cloud is bound. Note that the clouds used in all simulations here were unbound. Alternatively, the virial parameter $\alpha_\mathrm{vir}$ (\citealt{BMK92}; \citealt{MZ92}) takes into account that for a cloud to be in virial equilibrium it must have $\mathcal{T}/|W| = 0.5$. Thus calculating the value
\begin{equation} \label{eq:alpha_vir}
	\alpha_\mathrm{vir} \equiv \frac{2\mathcal{T}}{|W|}
\end{equation}
and finding $\alpha_\mathrm{vir} = 1$ implies virial equilibrium, while for boundedness $\alpha_\mathrm{vir} \le 2$. The value itself only differs from $\alpha$ by a factor of two. Note that in the literature these two parameters are often both given the symbol `$\alpha$', and so to differentiate them we use the `vir' subscript with the second. (See \citealt{BP06} for some of the virial parameter's shortcomings, including its neglecting to take into account that turbulence not only hinders star formation but can also help it through the formation of dense structure.)

\begin{table} \centering
	\caption{Star formation efficiency per free-fall time along with the ratio of kinetic plus thermal to gravitational energy, $\alpha$, and the virial parameter, $\alpha_\mathrm{vir}$. The values given for SFE$_\mathrm{ff}$ were calculated using Equation~\eqref{eq:sfeff} with $t_0 = 0.29 t_\mathrm{ff}$ and $t=1.38 t_\mathrm{ff}$ and are the data shown in Figure~\ref{fig:sfeff_vs_alpha}; the runs indicated with a dagger did not proceed that far, and so have values of SFE$_\mathrm{ff}$ given for the latest time they each reached.}
	\begin{tabular}{@{}lccccccc}
		\hline
		Run base name & Envelope? & $\alpha$ & $\alpha_\mathrm{vir}$ & SFE$_\mathrm{ff}$ \\
		\hline
		\multirow{2}{*}{3pc--medres} & X & 5.59 & 11.2 & 0.00 \\
		& E & 3.18 & 6.36 & 0.05 \\
		\hline
		\multirow{2}{*}{4pc--vlowres} & X & 3.15 & 6.31 & 0.05 \\
		& E & 2.28 & 4.56 & 0.18 \\
		\hline
		\multirow{2}{*}{4pc--lowres} & X & 3.15 & 6.31 & 0.05 \\
		& E & 2.28 & 4.56 & 0.17 \\
		\hline
		\multirow{2}{*}{4pc--medres} & X & 3.15 & 6.31 & 0.05 \\
		& E & 2.28 & 4.56 & 0.15 \\
		\hline
		\multirow{2}{*}{4pc--highres} & X$^\dagger$ & 3.15 & 6.31 & 0.03 \\
		& E$^\dagger$ & 2.28 & 4.56 & 0.08 \\
		\hline
		\multirow{2}{*}{5pc--vlowres} & X & 2.00 & 4.00 & 0.13 \\
		& E & 1.66 & 3.32 & 0.28 \\
		\hline
		\multirow{2}{*}{5pc--lowres} & X & 2.00 & 4.00 & 0.15 \\
		& E$^\dagger$ & 1.66 & 3.32 & 0.15 \\
		\hline
		\multirow{2}{*}{5pc--medres} & X & 2.00 & 4.00 & 0.14 \\
		& E & 1.66 & 3.32 & 0.25 \\
		\hline
		\multirow{2}{*}{5pc--highres} & X$^\dagger$ & 2.00 & 4.00 & 0.08 \\
		& E$^\dagger$ & 1.66 & 3.32 & 0.12 \\
		\hline
	\end{tabular}
	\label{tab:cloud_alphas}
\end{table}

Of interest is the star formation efficiency per free-fall time, $\mathrm{SFE_{ff}}$, which we calculated at time $t$ by
\begin{equation} \label{eq:sfeff}
	\mathrm{SFE}_\mathrm{ff}(t) = \frac{\mathrm{SFE}(t) - \mathrm{SFE}(t_0)}{t - t_0}
\end{equation}
(see e.g. \citealt{DEA14}). For $t_0$ we used the time at which the envelope was removed from the $5\,\mathrm{pc}$ cloud, $0.29t_\mathrm{ff}$. At this point structure had been generated and thus the clouds can be considered to have been in a star forming configuration. We calculated $\mathrm{SFE_{ff}}$ for each simulation at $t=1.38 t_\mathrm{ff}$. This was far enough into the runs that a reasonable fraction of mass had been allowed to convert to sink particles, but still early enough that data for most simulations was available. Furthermore, the period $t-t_0$ is almost exactly $2\,\mathrm{Myr}$, a stellar age often assumed in observational analysis of star formation (e.g. \citealt{EEA09}; \citealt{HEA10}; \citealt{PP13}). The values of $\mathrm{SFE_{ff}}$ are shown along with $\alpha$ and $\alpha_\mathrm{vir}$ for all simulations in Table~\ref{tab:cloud_alphas}. The runs which ended before $1.38 t_\mathrm{ff}$ are given with $\mathrm{SFE_{ff}}$ for the latest time available.

In Figure~\ref{fig:sfeff_vs_alpha} we show these data in a plot of $\mathrm{SFE_{ff}}$ against the virial parameter. As noted in Section~\ref{s:method}, the $5\,\mathrm{pc}$ cloud was the most bound. The cloud alone was set up to have twice as much kinetic as gravitational energy, and so is found with $\alpha = 2.00$ as the thermal energy in the cloud was negligible when compared to kinetic energy. While the envelope was warm and so provided a great amount of thermal energy, the extra mass provided even more gravitational energy, resulting in the overall effect of the envelope being increase the $\alpha$. This was the case with all runs.

With the shift in $\alpha$ provided by the envelope, it can be seen in Figure~\ref{fig:sfeff_vs_alpha} that, with some scatter, a relation exists between boundedness and $\mathrm{SFE_{ff}}$. This is particularly seen with E3pc--medres and the X4pc runs which lay nearly on top of one another despite there being a factor of $2.37$ difference in star forming cloud mass. X3pc--medres formed no stars and thus had $\mathrm{SFE_{ff}} = 0$. If one does form a relation between the plotted quantities, then it may be that $\mathrm{SFE_{ff}}$ would go to zero for all values of $\alpha$ above perhaps four or five. To demonstrably show this would however require more simulations to fill this region.

\subsection{Removing the envelope at later times}

\begin{figure} \centering
	\includegraphics[width=9cm]{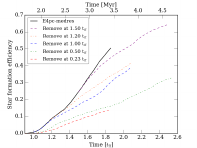}
	\caption{Star formation efficiency of E4pc--medres and all derivative simulations in which the envelope was removed at times later than $0.23 t_\mathrm{ff}$ (the original). Whether it took place before or after star formation had commenced, the loss of the envelope always caused a drop in the SFE. Those clouds which lost their envelopes afterwards typically followed E4pc--medres for a short while before peeling away.}
	\label{fig:sfe_r4pc_late_kill_linlin}
\end{figure}

It is apparent that removing the envelope early on led to very different star formation efficiencies. We decided to run an extra set of simulations derived from E4pc--medres in which the envelope was removed at several times before and after $1 t_\mathrm{ff}$, in an attempt to determine if there might be a critical point beyond which the removal no longer affected the SFE. The SFE is plotted against time for these simulations in Figure~\ref{fig:sfe_r4pc_late_kill_linlin}, which shows the original E4pc--medres simulation and others in which the envelope was removed at $0.23$ (the original X4pc--medres), $0.50$, $1.00$, $1.20$ and $1.50 t_\mathrm{ff}$.

Star formation in E4pc--medres commenced at $0.93 t_\mathrm{ff}$. Examination of Figure~\ref{fig:sfe_r4pc_late_kill_linlin} shows that the two simulations in which the envelope was removed before the first free fall time began forming stars only after that time. The simulations in which the envelope was removed from $1.00 t_\mathrm{ff}$ onwards generally followed the E4pc--medres curve for a short period of at most $\approx 0.2 t_\mathrm{ff}$ after the split and then fell away after a brief period during which the cloud reacted to the sudden loss of the envelope. This was not immediate: the gradients of the SFE in `late loss' ($\ge 1.00 t_\mathrm{ff}$) runs began high, and slowly tended down over time towards values more closely resembling the gradients of the two `early loss' ($< 1.00 t_\mathrm{ff}$) runs. This can in particular be seen for the two longest running simulations, those which lost their envelopes at $0.50 t_\mathrm{ff}$ and $1.50 t_\mathrm{ff}$. It is possible however that the downturn in the SFE as it approached $0.7$ in the latter run actually occurred due to star forming material becoming more scarce.

Figure~\ref{fig:sfe_r4pc_late_kill_linlin} makes it clear that a removal of the envelope inevitably led to a reduction in the SFE, though a later removal brings it closer to that of the simulation which retained it. Thus no critical point for its effect would appear to exist, at least within the simulation runtimes achieved.

\section{Conclusions} \label{s:conclusions}

Star formation principally occurs in the cold and dense gas of giant molecular clouds. Yet these clouds will themselves be enshrouded with a warm atomic envelope, which can help to bind the cloud and thus keep it contained. It is possible however that feedback from nearby massive stars, whether in the form of ionisation, winds or supernovae, could blow away the envelope. As the envelope alters the cloud's boundedness, we asked how its loss would change the star formation efficiency of the remaining dense molecular cloud.

Using smoothed particle hydrodynamics we ran isothermal star forming simulations of three dense, cold and turbulent clouds, initially spherical, with radii of $3$, $4$ and $5\,\mathrm{pc}$. The smaller clouds were themselves derived from the 5 pc cloud. In isolation all three clouds would have been unbound with $\alpha$, the ratio of the kinetic plus thermal to gravitational energy, of $5.59$, $3.15$ and $2.00$ respectively. The clouds were not evolved in isolation however, but rather they were placed within a warm envelope which provided a confining pressure boundary. The envelopes had a density a factor of ten lower, extending outwards in all cases to $10\,\mathrm{pc}$. With their presence the values of $\alpha$ were lowered to $3.18$, $2.28$ and $1.66$, again respectively with the $3$, $4$ and $5\,\mathrm{pc}$ clouds.

These simulations were evolved through to the point where stars could form, numerically represented by sink particles. However, a parallel simulation was created from each of the originals after $0.25 t_\mathrm{cross}$ of evolution. In these runs, the warm envelope was instantaneously removed to leave only the cold central cloud. This was our first order approximation to the envelope's removal by feedback.

We found for all three clouds that the removal of the envelope led to drastically lower levels of structure forming within the cloud while the external layers expanded. This naturally also caused the average densities to fall. As such it became harder for gas to become self gravitating and the levels of star formation fell.

The star formation efficiencies (SFEs) were calculated as a function of time. For all pairs of simulations (with and without the envelope), those which retained the envelope showed higher levels of star formation. At a fiducial time of $1.38 t_\mathrm{ff}$, the $4$ and $5\,\mathrm{pc}$ clouds retaining the envelope showed SFEs higher by factors of 1.7 to 2.7 compared to the runs in which it was lost. The $4\,\mathrm{pc}$ simulations ran until a much later time and by $\approx 1.8 t_\mathrm{ff}$ the discrepancy was even higher with SFEs of $\approx 0.5$ falling to $\approx 0.15$ with the loss of the envelope.

The $3\,\mathrm{pc}$ cloud was the most unbound.  At the same time of $1.38t_\mathrm{ff}$, it had an SFE of $5.67\times 10^{-2}$ when the envelope was present. In its counterpart which lost the envelope, no star formation whatsoever took place. 

Considering the star formation efficiencies per free-fall time (SFE$_\mathrm{ff}$) and plotting these against $\alpha$ shows the simulations to be nearly a full sequence, with the $3\,\mathrm{pc}$ cloud with an envelope falling on almost the same point as the $4\,\mathrm{pc}$ cloud lacking an envelope, despite the larger cloud having $2.37$ times the mass of the smaller.

We also ran a set of simulations derived from the $4\,\mathrm{pc}$ cloud in which the envelope was removed at later times. Even when it was lost as late as $1.50 t_\mathrm{ff}$, the SFE fell away to a slope resembling those simulations which had lost the envelope at an earlier stage. In the runs in which the envelope was removed after star formation had begun, the SFE only responded to the loss after a short period of time of at most $\approx 0.2 t_\mathrm{ff}$.

A more realistic form of feedback may have brought about less extreme changes in the SFE than those reported here. For example, the envelope could have been lost only partially across some small solid angle, leading perhaps to the cloud's expansion through the new opening as a cold outflow. Or, the envelope could have been lost slowly rather than removed instantaneously. However, the end state would have still resembled one of the simulations in which the envelope was removed at a later time. This implies that even then a drop in the SFE would be inevitable.

\section*{Acknowledgements}

The authors thank Paul Clark for his contribution towards the original concept of this paper and the anonymous referee for their helpful comments. Our column density plots were produced using Daniel Price's \textsc{splash} software \citep{PR07}. The authors gratefully acknowledge support from the ECOGAL project, grant agreement 291227, funded by the European Research Council under ERC-2011-ADG. This work used the compute resources of the St Andrews MHD Cluster. This work used the DiRAC Complexity system, operated by the University of Leicester IT Services, which forms part of the STFC DiRAC HPC Facility (www.dirac.ac.uk). This equipment is funded by BIS National E-Infrastructure capital grant ST/K000373/1 and STFC DiRAC Operations grant ST/K0003259/1. DiRAC is part of the National E-Infrastructure.


\begin{thebibliography}{}

\bibitem[\protect\citeauthoryear{Abgrall et al.}{1992}]{AEA92}Abgrall H., Le Bourlot J., Pineau des For\^{e}ts G., Roueff E., Flower D. R., Heck L., 1992, A\&A, 253, 525

\bibitem[\protect\citeauthoryear{Ballesteros-Paredes}{2006}]{BP06}Ballesteros-Paredes J., 2006, MNRAS, 372, 443

\bibitem[\protect\citeauthoryear{Bate, Bonnell \& Price}{Bate et al.}{1995}]{BBP95}Bate M. R., Bonnell I. A., Price N. M., 1995, MNRAS, 227, 362
  
\bibitem[\protect\citeauthoryear{Benz}{1990}]{BE90}Benz W., 1990, in  Buchler R. J., ed., Proceedings of the NATO Advanced Research Workshop on The Numerical Modelling of Nonlinear Stellar Pulsations: Problems and Prospects, Kluwer Academic Publishers, Dordrecht, p. 269
  
\bibitem[\protect\citeauthoryear{Bertoldi \& McKee}{1992}]{BMK92}Bertoldi F., McKee C. F., 1992, ApJ, 395, 140
  
\bibitem[\protect\citeauthoryear{Bolatto et al.}{2008}]{BEA08}Bolatto A. D., Leroy A. K., Rosolowsky E., Walter F., Blitz L., 2008, ApJ, 686, 948

\bibitem[\protect\citeauthoryear{Bonnell, Bate \& Vine}{Bonnell et al.}{2003}]{BBV03}Bonnell I. A., Bate M. R., Vine S.G., 2003, MNRAS, 343, 413

\bibitem[\protect\citeauthoryear{Bonnell, Larson \& Zinnecker}{Bonnell et al.}{2007}]{BLZ07}Bonnell I. A., Larson R. B., Zinnecker H., in Reipurth B., Jewitt D., Keil K., eds, Protostars and Planets V, University of Arizona Press, Tucson, p. 149
  
\bibitem[\protect\citeauthoryear{Bonnell et al.}{2011}]{BEA11}Bonnell I. A., Smith R. J., Clark P. C., Bate M. R., 2011, MNRAS, 410, 2339

\bibitem[\protect\citeauthoryear{Boss \& Bodenheimer}{1979}]{BB79}Boss A. P., Bodenheimer P., 1979, ApJ, 234, 289

\bibitem[\protect\citeauthoryear{Clark, Bonnell \& Klessen}{Clark et al.}{2008}]{CBK08}Clark P. C., Bonnell I. A., Klessen R. K., 2008, MNRAS, 386, 3

\bibitem[\protect\citeauthoryear{Dale, Bonnell \& Whitworth}{Dale et al.}{2007}]{DBW07}Dale J. E., Bonnell I. A., Whitworth A. P., 2007, MNRAS, 375, 1291

\bibitem[\protect\citeauthoryear{Dale et al.}{2014}]{DEA14}Dale J. E., Ngoumou J., Ercolano B., Bonnell I. A., 2014, MNRAS, 442, 694

\bibitem[\protect\citeauthoryear{Dobbs, Bonnell \& Clark}{Dobbs et al.}{2005}]{DBC05}Dobbs C. L., Bonnell I. A., Clark P. C., 2005, MNRAS, 360, 2

\bibitem[\protect\citeauthoryear{Dobbs, Burkert \& Pringle}{Dobbs et al.}{2011}]{DBP11}Dobbs C. L., Burkert A., Pringle J. E., 2011, MNRAS, 413, 2935

\bibitem[\protect\citeauthoryear{Dobbs et al.}{2014}]{DEA142}Dobbs C. L. et al., 2014, in Beuther H. et al., eds, Protostars and Planets VI, University of Arizona Press, Tucson, p. 3

\bibitem[\protect\citeauthoryear{Elmegreen \& Lada}{1977}]{EL77}Elmegreen B. G., Lada C. J., 1977, ApJ, 214, 725

\bibitem[\protect\citeauthoryear{Evans et al.}{1999}]{E99}Evans N. J. II, 1999, ARA\&A, 37, 311
  
\bibitem[\protect\citeauthoryear{Evans et al.}{2009}]{EEA09}Evans N. J. II et al., 2009, ApJS, 181, 321

\bibitem[\protect\citeauthoryear{Federrath}{2013}]{F13}Federrath C., 2013, MNRAS, 436, 3167

\bibitem[\protect\citeauthoryear{Goldsmith \& Li}{2005}]{GL05}Goldsmith P. F., Li D., 2005, ApJ, 622, 938

\bibitem[\protect\citeauthoryear{Heiderman et al.}{2010}]{HEA10}Heiderman A. Evans N. J. II, Allen L. E., Huard T., Heyer M., 2010, ApJ, 723, 1019

\bibitem[\protect\citeauthoryear{Heyer \& Brunt}{2004}]{HB04}Heyer M. H., Brunt C. M., 2004, ApJ, 615, L45

\bibitem[\protect\citeauthoryear{Heyer et al.}{2009}]{HEA09}Heyer M., Krawczyk C., Duval J., Jackson J. M., 2009, ApJ, 699, 1092

\bibitem[\protect\citeauthoryear{Heyer et al.}{2016}]{HEA16}Heyer M., Gutermuth R., Urquhart J. S., Csengeri T., Wienen M., Leurini S., Menten K., Wyrowski F., 2016, A\&A, 588, A29

\bibitem[\protect\citeauthoryear{Kennicutt \& Evans}{2012}]{KE12}Kennicutt R. C., Evans N. J. II, 2012, ARA\&A, 50, 531

\bibitem[\protect\citeauthoryear{Krumholz \& Tan}{2007}]{KT07}Krumholz M. R., Tan J. C., 2007, ApJ, 654, 304

\bibitem[\protect\citeauthoryear{Krumholz, Dekel \& McKee}{Krumholz et al.}{2012}]{KDM12}Krumholz M. R., Dekel A., McKee C. F., 2012, ApJ, 745, 69

\bibitem[\protect\citeauthoryear{Krumholz}{2014}]{KR14}Krumholz M. R., Phys. Rep., 539, 49
  
\bibitem[\protect\citeauthoryear{Lada \& Lada}{2003}]{LL03} Lada C. J., Lada E. A., 2003, ARA\&A, 41, 57

\bibitem[\protect\citeauthoryear{Larson}{1981}]{L81}Larson R. B., 1981, MNRAS, 194, 809

\bibitem[\protect\citeauthoryear{Larson}{2005}]{LA05}Larson R. B., 2005, MNRAS, 359, 211

\bibitem[\protect\citeauthoryear{Le Bourlot et al.}{1993}]{LBEA93}Le Bourlot J., Pineau des For\^{e}ts G., Roueff E., Flower D. R., 1993, A\&A, 267, 233

\bibitem[\protect\citeauthoryear{Louvet et al.}{2014}]{LEA14}Louvet F., et al., 2014, A\&A, 570, 15

\bibitem[\protect\citeauthoryear{Mac Low \& Klessen}{2004}]{MLK04}Mac Low M.-M., Klessen R. S., 2004, Rev. Mod. Phys., 76, 125

\bibitem[\protect\citeauthoryear{McKee \& Zweibel}{1992}]{MZ92}McKee C. F., Zweibel E. G., 1992, ApJ, 399, 551

\bibitem[\protect\citeauthoryear{MacLachlan et al.}{2015}]{MLEA15}MacLachlan J. M., Bonnell I. A., Wood K., Dale J. E., 2015, A\&A, 573, 112

\bibitem[\protect\citeauthoryear{Monaghan}{1992}]{MO92}Monaghan J.J., 1992, ARA\&A, 30, 543

\bibitem[\protect\citeauthoryear{Padoan \& Nordlund}{2011}]{PN11}Padoan P., Nordlund \AA., 2011. ApJ, 730, 40

\bibitem[\protect\citeauthoryear{Parmentier \& Pfalzner}{2013}]{PP13}Parmentier G., Pfalzner S., 2013, A\&A, 549, A132

\bibitem[\protect\citeauthoryear{Price}{2007}]{PR07}Price D. J., 2007, PASA, 24, 159

\bibitem[\protect\citeauthoryear{Price \& Bate}{2009}]{PB09}Price D. J., Bate M. R., 2009, MNRAS, 398, 33

\bibitem[\protect\citeauthoryear{Price \& Monaghan}{2004}]{PM04}Price D.J., Monaghan J.J., 2004, MNRAS, 348, 139
  
\bibitem[\protect\citeauthoryear{Price \& Monaghan}{2007}]{PM07}Price D. J., Monaghan J. J., 2007, MNRAS, 374, 1347

\bibitem[\protect\citeauthoryear{Rosolowsky et al.}{2003}]{REA03}Rosolowsky E., Engargiola G., Plambeck R., Blitz L., 2003, ApJ, 599, 258

\bibitem[\protect\citeauthoryear{Salim, Federrath \& Kewley}{Salim et al.}{2015}]{SFK15}Salim D. M., Federrath C., Kewley L. J., 2015, ApJ, 806, L36

\bibitem[\protect\citeauthoryear{Shu, Adams \& Lizano}{Shu et al.}{1987}]{SAL87}Shu F. H., Adams F. C., Lizano S., 1987, ARA\&A, 25, 23

\bibitem[\protect\citeauthoryear{Tohline}{1980}]{T80}Tohline J. E., 1980, ApJ, 235, 866

\bibitem[\protect\citeauthoryear{Wong et al.}{2011}]{WEA11}Wong T. et al., 2011, ApJS, 197, 16

\end{thebibliography}
\end{document}